\documentclass[useAMS,usenatbib]{mn2e}

\usepackage{graphicx}
\title[Abell~70]{A barium central star binary in the Type-I diamond ring planetary nebula Abell~70\thanks{Based on observations made with Gemini South under program GS-2009A-Q-35 and the Very Large Telescope at Paranal Observatory under programs 083.D-0654(A) and 085.D-0629(A).}}
\author[B. Miszalski et al.]{B. Miszalski,$^{1,2,3}$\thanks{E-mail: brent@saao.ac.za} H. M. J. Boffin,$^{4}$ D. J. Frew,$^{5}$ A. Acker,$^{6}$ J. K\"oppen,$^{6,7,8}$\newauthor A. F. J. Moffat$^{9}$ and Q. A. Parker$^{5,10}$\\
$^{1}$South African Astronomical Observatory, PO Box 9, Observatory, 7935, South Africa\\
$^{2}$Southern African Large Telescope Foundation, PO Box 9, Observatory, 7935, South Africa\\
$^{3}$Centre for Astrophysics Research, STRI, University of Hertfordshire, College Lane Campus, Hatfield AL10 9AB\\
$^{4}$European Southern Observatory, Alonso de Cordova 3107, Casilla 19001, Santiago, Chile\\
$^{5}$Department of Physics and Astronomy, Macquarie University, Sydney, NSW 2109, Australia\\
$^{6}$Observatoire astronomique de Strasbourg, Universit\'e de Strasbourg, CNRS, UMR 7550, 11 rue de l'Universit\'e, F-67000 Strasbourg, France\\
$^{7}$International Space University, Parc d'Innovation, 1 Rue Jean-Dominique Cassini, F-67400, Illkirch-Graffenstaden, France\\
$^{8}$Institut f\"ur Theorestische Physik und Astrophysik, Universit\"at Kiel, D-24098, Kiel, Germany\\
$^{9}$D\'ept. de physique, Univ. de Montr\'eal C.P. 6128, Succ. Centre-Ville, Montr\'eal, QC H3C 3J7, and Centre de recherche\\ en astrophysique du Qu\'ebec, Canada\\
$^{10}$Australian Astronomical Observatory, Epping, NSW 1710, Australia
}
\begin{document}

\date{Accepted Received ; in original form }

\pagerange{\pageref{firstpage}--\pageref{lastpage}} \pubyear{2002}

\maketitle

\label{firstpage}

\begin{abstract}
   Abell~70 (PN G038.1$-$25.4, hereafter A~70) is a planetary nebula (PN) known for its diamond ring appearance due a superposition with a background galaxy. The previously unstudied central star is found to be a binary consisting of a G8IV-V secondary at optical wavelengths and a hot white dwarf (WD) at UV wavelengths. The secondary shows Ba~II and Sr~II features enhanced for its spectral type that, combined with the chromospheric H$\alpha$ emission and possible 20--30 km/s radial velocity amplitude, firmly classifies the binary as a Barium star. The proposed origin of Barium stars is intimately linked to PNe whereby wind accretion pollutes the companion with dredged-up material rich in carbon and s-process elements when the primary is experiencing thermal pulses on the Asymptotic Giant Branch (AGB). A~70 provides further evidence for this scenario together with the other very few examples of Barium central stars. The nebula is found to have Type-I chemical abundances with helium and nitrogen enrichment, which when combined with future abundance studies of the central star, will establish A~70 as a unique laboratory for studying s-process AGB nucleosynthesis. We also discuss guidelines to discover more binary central stars with cool secondaries in large orbits that are required to balance our knowledge of binarity in PNe against the currently better studied post common-envelope binary central stars.
\end{abstract}

\begin{keywords}
   planetary nebulae: individual: PN G038.1-25.4 - planetary nebulae: general - stars: chemically peculiar - stars: AGB and post-AGB - binaries: general - binaries: symbiotic
\end{keywords}

   \section{Introduction}
Planetary nebulae (PNe) are the ionised nebulae ejected by low-intermediate mass stars that have undergone extensive mass loss during the AGB phase. The central stars of planetary nebulae (CSPN) constitute a rich resource to study the late stages of binary stellar evolution. At least 40 close binary CSPN are known that have orbital periods less than $\sim$1 day (Miszalski et al. 2011a) and these make up at least $17\pm5$\% of all CSPN (Miszalski et al. 2009a). With their short-lived nebulae ($\sim$10$^4$ yrs) close binary CSPN are assured to have just recently passed through the common-envelope (CE) phase (Iben \& Livio 1993). With significantly less time to undergo further angular momentum loss compared to other more evolved post-CE binaries (Schreiber \& G\"ansicke 2003), the orbital periods of close binary CSPN reflect the true post-CE distribution. This makes them an excellent but relatively unexplored population to constrain CE population synthesis models. 

Miszalski et al. (2009a) showed that there is a drop-off in the post-CE period distribution for periods $\ga$ 1 day consistent with more evolved post-CE binaries (Rebassa-Mansergas et al. 2008). While the cause behind this sharp drop-off is not yet understood (Davis et al. 2010), even less is known about intermediate period binaries in PNe ($\sim$100--1500 days) that are predicted by some models (Yungelson et al. 1993; Han et al. 1995; De Marco et al. 2009) and their anticipated connection with the period distribution of post-AGB binaries that precede PNe (Van Winckel 2003a). At present there is essentially no hard evidence to support the existence of these binaries in PNe, primarily because there have been no substantial long-term radial velocity monitoring campaigns. This is starting to change with some bright central stars included in the survey described by Van Winckel et al. (2010).

The best way to find intermediate period binaries is to look for giant or sub-giant companions whose orbits must be sufficiently large to accommodate their larger radii. As such cool companions are typically more luminous than their white dwarf (WD) companions, these binaries can usually only be found with the aid of UV photometry (e.g. Maxted et al. 2009). The arduous UV selection is already done for PNe where the nebula acts as a natural signpost for the presence of a hot WD. All that is left is to identify a suitable CSPN candidate too cool to ionise the PN and prove it has a physical connection to the nebula. A small number of CSPN have suspected cool central stars (e.g. Lutz 1977; Tab. 4 of De Marco 2009), but very few have been studied in sufficient detail to rule out a line-of-sight superposition (excluding K~1-6, see Frew et al. 2011). Poor UV sensitivity is a likely explanation for suggestions that some are single CSPN caught momentarily during a born-again phase (Bond \& Pollacco 2002). 

Perhaps the most studied of the cool central stars are those of the so-called A~35 type (Bond, Ciardullo \& Meakes 1993). The initial list included A~35, LoTr~1 and LoTr~5 which have rapidly rotating sub-giants or giants accompanied by very hot white dwarfs peaking at UV wavelengths ($T_\mathrm{eff}\ga100$ kK). The most interesting aspect of these binaries is that the secondaries of A~35 and LoTr~5 both exhibit enhanced barium abundances (Th\'evenin \& Jasniewicz 1997), while this has yet to be demonstrated for LoTr~1. Since these initial discoveries Bond, Pollacco \& Webbink (2003) added WeBo~1 to the list and Frew (2008) raised suspicions that the nebula of A~35 may not be a bona-fide PN.\footnote{Note however that A~35 is likely to have passed through a PN phase at some stage to create the stellar abundances currently observed.} Further additions of HD~330036 and AS~201 may be drawn from barium enhanced yellow or D'-type symbiotic stars which show extended nebulae (Schmid \& Nussbaumer 1993; Pereira et al. 2005), however the classification of such objects as PNe is controversial (Corradi 2003; Jorissen et al. 2005). All these binaries may be classified as Barium stars, peculiar red giants characterised by an overabundance of carbon and s-process heavy elements such as barium and strontium (Bidelman \& Keenan 1951). They make up $\sim$1\% of all G--K giants, all of them are binaries with apparently WD companions (McClure et al. 1980) and have orbital periods ranging from 87 up to $\sim$2000 days (Jorissen \& Boffin 1992; McClure \& Woodsworth 1990). A smaller sample of Barium dwarfs are known and are thought to evolved into Barium giants, although they are less studied on the whole (e.g. North, Jorissen \& Mayor 2000; Gray et al. 2011). 

Boffin \& Jorissen (1988) devised what is now believed to be the canonical model for the formation of these stars. In this scenario the wind of its companion pollutes the future barium-enhanced star, most likely while it is on the main-sequence. This wind transfers carbon and s-process elements dredged up during thermal pulses on the AGB. The AGB star then evolved into a WD, which may have produced a PN, while the contaminated star evolved to become a red giant with chemical anomalies: a Barium star. This accounts for the few known PNe caught exactly during this phase, but there remains much work to identify and characterise further examples.  None of the three firm examples, A~35, LoTr~5 and WeBo~1, have determined orbital periods, while photometric monitoring has revealed the cool components to be rapid rotators with rotation periods of a few days. The rapid rotation could possibly have been caused by mass accretion (Jeffries \& Stevens 1996; Theuns et al. 1996).

In this work we present UV and optical observations of the \emph{diamond ring}\footnote{The \emph{diamond ring} effect arises from the superposition of the background galaxy 6dFGS gJ203133.1-070502 (Jones et al. 2004) with the northern edge of the ring-shaped nebula.} PN A~70 (PN G038.1$-$25.4, Abell 1966) that prove the existence of a Barium star binary CSPN. The 2 micron all sky survey (2MASS, Skrutskie et al. 2006) recorded an unusually red $J-H=+0.70$ mag colour for the CSPN which suggested a sub-giant or giant classification. 
This prompted our investigation of the object even though it may have been possible to reach this conclusion from the original photographic magnitudes on blue (18.6 mag) and red (17.4 mag) plates. More recent studies of the Abell sample lacked colour information for the CSPN of A 70 and mostly repeated earlier measurements (Kaler 1983; Kaler \& Jacoby 1989; Kaler, Shaw \& Kwitter 1990). Narrow-band imaging of the nebula was acquired by Jewitt, Danielson \& Kupferman (1986), Schwarz, Corradi \& Melnick (1992) and most notably Hua, Dopita \& Martinis (1998). 

\section{The Binary Central Star}
\subsection{Observations}
\label{sec:obs}
We obtained spectroscopic observations of A~70 with Gemini South under program GS-2009A-Q-35 and the VLT under programs 083.D-0654(A) and 085.D-0629(A). Table \ref{tab:obs} summarises the 0.7\arcsec-wide longslit observations made using the Gemini Multi-Object Spectrograph (GMOS, Hook et al. 2004) and the focal reducer and low dispersion spectrograph (FORS2, Appenzeller et al. 1998) that had the blue-optimised E2V detector installed. Basic data reduction was performed using the Gemini \textsc{iraf} package and the ESO FORS pipeline after which the \textsc{iraf} task \textsc{apall} was used to trace and extract one-dimensional spectra. In all cases the surrounding nebula emission was subtracted close to the central star. Flux calibration was also applied using spectrophotometric standard stars observed during the respective programs in the usual fashion. The S/N reached in the continuum near $\lambda5130$ \AA\ was measured using the \textsc{splot} task in \textsc{iraf} to range between 19 (FORS2 2009) and 30 (GMOS B). Some narrow-band images were also observed by GMOS (see Sect. \ref{sec:imaging}). 

\begin{table*}
   \centering
   \caption{Summary of Gemini South and VLT observations.}
   \label{tab:obs}
   \begin{tabular}{lclcccrc}
      \hline\hline
      Spectrum & MJD & Grating & $\lambda$ & Resolution  & Dispersion & Position Angle & Exptime\\
               &     &           &   (\AA)      &    (FWHM, \AA) & (\AA/pix) & ($^\circ$) &  (s) \\
      \hline
      GMOS B & 54944.83 & B1200 & 4085--5550 & 1.6 & 0.23 & 90 & 1800 \\
      GMOS R & 54944.85 & R400 & 4650--8890 & 5.3 & 0.68 & 90 & 1800 \\
      FORS2 2009$^\dag$ & 55026.40 & 1200g & 4088--5555 & 1.5 & 0.72 & 107 & 2400\\
      FORS2 2010A & 55364.41 & 1200g & 4088--5559 & 1.5 & 0.36 & 108 & 1800 \\
      FORS2 2010B & 55366.40 & 1200g & 4088--5559 & 1.5 & 0.36 & 87 & 1800 \\
      \hline
   \end{tabular}
   \begin{flushleft}
      $^\dag$ Affected by clouds.
   \end{flushleft}
\end{table*}

\subsection{An s-process enhanced G8IV-V companion}
\begin{figure}
   \begin{center}
      \includegraphics[scale=0.37,angle=270]{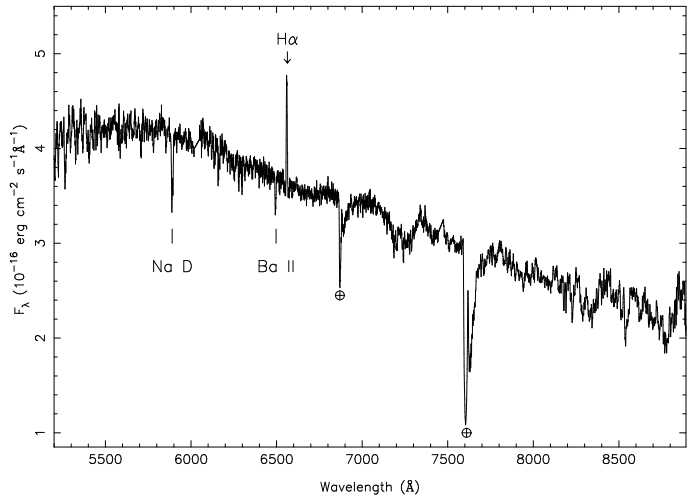}
   \end{center}
   \caption{Dereddened GMOS R spectrum of A~70. }
   \label{fig:red}
\end{figure}

Figure \ref{fig:red} shows the GMOS R spectrum dereddened with $c(H\beta)=0.07$ mag (Acker et al. 1992). The He~II and H~I absorption lines expected for a typical hot WD are absent. Instead, we find only a cool star that validates our 2MASS colour selection. In order to determine its spectral type we first convolved Johnson $V$ and Cousins $R$ filters with the flux-calibrated spectrum to obtain Vega magnitudes using the \textsc{synphot} \textsc{iraf} package. Table \ref{tab:cool} lists these magnitudes in addition to other magnitudes from the literature. The $(V-R_C)_0=0.45$ mag intrinsic colour of the star was compared against the same colour measured from Pickles (1998) spectra in the same way to find a best match of G8IV-V (0.46/0.43). The agreement is reasonable for now until improved photometric and spectroscopic observations are made. The luminosity class of the G8IV-V component is discussed in Sect. \ref{sec:distance} and the WD is discussed in Sect. \ref{sec:hot}.

\begin{table}
   \centering
   \caption{Observed ($m$) and dereddened ($m_0$) magnitudes of the G8IV-V component of the CSPN.} 
   \begin{tabular}{crrc}
      \hline\hline
      Waveband & $m$ & $m_0$ & Reference\\
      \hline
      $B$         & 18.6 & 18.3 & Abell (1966)\\
      $R$          & 17.4 &17.3 & Abell (1966)\\       
      Johnson $V$ & 17.82 & 17.67  & this work\\
      Cousins $R$  & 17.33 & 17.22  & this work\\
      Gunn $i$   & 16.85 & 16.77  & Epchtein et al. (1999) \\ 
      $J$   & 16.16 & 16.12       & Skrutskie et al. (2006)\\
      $H$   & 15.46 & 15.43       & Skrutskie et al. (2006)\\
      $K_s$   & 15.30 & 15.29     & Skrutskie et al. (2006)\\
      \hline
   \end{tabular}
   \label{tab:cool}
\end{table}

\begin{figure}
   \begin{center}
      \includegraphics[scale=0.38,angle=270]{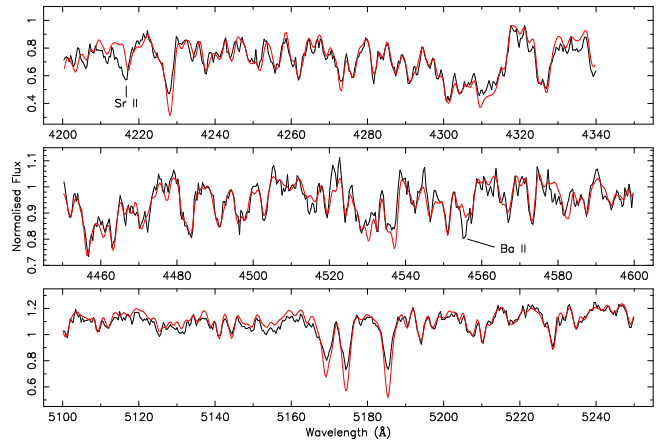}\\
   \end{center}
   \caption{Normalised VLT FORS2 2010B spectrum (black) overlaid with the G8 IV-V comparison star HD24616 (red, Bagnulo et al. 2003). Note the absence of strong Sr II $\lambda$4216 and Ba II $\lambda$4554 in HD25616, whereas they are both enhanced in A~70.}
   \label{fig:zoom}
\end{figure}

To look for s-process enhancement we compared the FORS2 2010B spectrum against the G8IV-V star HD24616 as observed by the UVES Paranal Observatory Project (Bagnulo et al. 2003). Figure \ref{fig:zoom} shows the FORS2 2010B spectrum overplotted with the HD24616 spectrum (both rebinned to a resolution of FWHM=0.75 \AA). As both Sr~II $\lambda$4216 and Ba~II $\lambda$4554 are enhanced in A~70 compared to the reference star this is a real over-abundance rather than a luminosity effect. This result firmly places A~70 amongst the three other Barium star CSPN of A~35, LoTr~5 and WeBo~1 (Th\'evenin \& Jasniewicz 1997; Bond et al. 2003). Except for the problematic case of A~35 (Frew 2008), the others all have Ba giants, the Ba dwarf or sub-giant in A~70 provides even greater support for the standard model of Ba star formation (e.g. Gray et al. 2011). A Ba1 sub-type is consistent with the weak Ba~II $\lambda$4554 line, present but weak carbon features (CN, CH and C$_2$) and the relatively early spectral type (L\"u et al. 1983). Later spectral types generally have stronger features (see e.g. Bond et al. 2003). A meaningful stellar abundance analysis is left to future work once high resolution spectra are obtained, however a preliminary analysis of the FORS2 spectra suggests a [Ba/Fe] overabundance of $\sim$0.5 dex. 

\subsection{Chromospheric H$\alpha$ emission}
\begin{figure}
   \begin{center}
      \includegraphics[angle=270,scale=0.35]{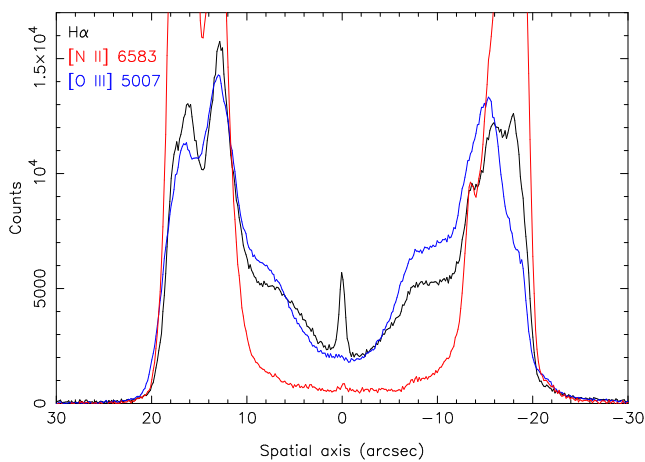}
   \end{center}
   \caption{Stellar continuum subtracted traces of key emission lines along the spatial axis of the GMOS R spectrum. The residual H$\alpha$ emission is likely to be chromospheric as seen in other Barium star CSPN. 
   }
   \label{fig:trace}
\end{figure}

A notable feature in Fig. \ref{fig:red} is the residual H$\alpha$ emission. The spatial resolution of the GMOS R spectrum (0.145 \arcsec/pixel) allowed for the nebula to be accurately subtracted from small windows either side of the CSPN. Figure \ref{fig:trace} verifies that the emission is real with traces of the H$\alpha$, [N~II] $\lambda$6584 and [O~III] $\lambda$5007 emission lines along the spatial direction. The stellar contribution was removed with adjacent traces of the same width (6.8 \AA). Excess emission remains only in the H$\alpha$ line leaving no doubt that it is real. The origin is chromospheric in nature and probably originates from a hot spot on the cool star (Th\'evenin \& Jasniewicz 1997). Stellar H$\alpha$ emission is also seen in A 35 where its radial velocity variations seem to follow the rotational period of the cool star as derived from photometric observations (Acker \& Jasniewicz 1990) and chromospheric Ca~II H emission is seen in WeBo~1 (Bond et al. 2003). 

\subsection{Radial velocities}
\label{sec:rv}
The higher resolution blue spectra in Tab. \ref{tab:obs} are well-suited to determine whether the radial velocity (RV) of the nebula matches the G8IV-V star and to probe variations that may be due to orbital motion. As the GMOS B spectrum was the only one observed with a contemporaneous arc lamp exposure it served as our reference spectrum. We re-extracted all spectra to contain both the CSPN and nebula emission so that features from both components could be measured from the same spectrum. From both [O~III] emission lines in the GMOS B spectrum we measured a heliocentric nebula radial velocity of $V_\mathrm{neb}=-72\pm3$ km/s with the \textsc{rvsao} task \textsc{emsao} (Kurtz \& Mink 1998). This value is in good agreement with $-79\pm18$ km/s found by Meatheringham et al. (1998). The radial velocities of the [O~III] emission lines were also measured in the FORS2 spectra, which were then used to shift the FORS2 spectra to the same wavelength scale as the GMOS B spectrum. The H$\beta$ line was excluded from this process given its weakness and the potential influence of H$\beta$ absorption from the G8IV-V star on its measured velocity. 

Table \ref{tab:rv} lists the radial velocities of the G8IV-V star as measured from the mean velocity of three Mg~I absorption lines $\lambda$5167, $\lambda$5172 and $\lambda$5183 \AA. The errors are the standard deviation of the three measurements with the largest error expectedly belonging to the lowest S/N FORS2 2009 spectrum. Within the errors, the FORS2 data were consistent with $V_\mathrm{neb}$, proving that the G8IV-V star is physically connected to the nebula. The 24 km/s difference between $V_\mathrm{neb}$ and the GMOS B measurement was originally suspected to arise from the fact that the Mg~I absorption lines lie on a different CCD to the [O~III] emission lines, however a careful manual re-reduction of the data per CCD proved this to not be the case. If the one discrepant measurement is indeed true, then it would be consistent with typical Barium star amplitudes of 20--30 km/s (McClure \& Woodsworth 1990). Such an amplitude may be rather high for what should be a low-inclination object (Sect. \ref{sec:imaging}), so we consider the amplitude to be an upper limit. The apparent constancy of the FORS2 measurements is consistent with a few hundred day orbital period and the most suitable progenitor for the system would be a post-AGB binary (Van Winckel 2003a, 2003b).

\begin{table}
   \centering
   \caption{Radial velocities of the G8IV-V star.}
   \label{tab:rv}
   \begin{tabular}{lcl}
      \hline\hline
      Spectrum & MJD & HRV (km/s) \\ 
      \hline
      GMOS B & 54944.83 &  $-96\pm2$\\
      FORS2 2009 & 55026.40   & $-67\pm18$\\
      FORS2 2010A & 55364.41 & $-69\pm8$\\
      FORS2 2010B & 55366.40 &$-70\pm8$\\
      \hline
   \end{tabular}
\end{table}

\subsection{Detection of the WD and spectral energy distribution}
\label{sec:hot}
To rule out the possibility that the G8IV-V star is a single CSPN caught momentarily during a `born-again' phase (Bond \& Pollacco 2002), the WD of A~70 must be detected.
Fortunately, A~70 lies at a high enough Galactic latitude to be covered by the All Sky Imaging Survey (AIS) of \emph{GALEX} (Martin et al. 2005; Morrissey et al. 2007). The far-ultraviolet (FUV, $\sim$154 nm) and near-ultraviolet (NUV, $\sim$232 nm) images (Sect. \ref{sec:imaging}) have corresponding pipeline AB magnitudes of $16.48\pm0.04$ and $16.63\pm0.03$, respectively. These were however unsuitable since they included extended nebula emission. We performed aperture photometry on the images with a sky aperture radius incorporating most of the nebula to find FUV=$18.00\pm0.05$ mag and NUV=$18.9\pm0.1$ mag. Table \ref{tab:hot} lists the \emph{GALEX} AB magnitudes alongside estimated Vega-based magnitudes of the WD. These magnitudes were calculated by convolving a spectrum of NGC~7293 (Oke 1990), scaled to the \emph{GALEX} magnitudes, through the filters listed using \textsc{synphot}. Values of $A(NUV)=2A(V)$ and $A(FUV)=2.3A(V)$ were inferred from Fig. 4 of Cardelli et al. (1989) to deredden the \emph{GALEX} photometry, while standard waveband specific corrections were used at other wavelengths (Cardelli et al. 1989). As before $c(H\beta)=0.07$ mag or $A(V)=0.15$ mag was used (Acker et al. 1992).

\begin{figure}
   \begin{center}
      \includegraphics[angle=270,scale=0.37]{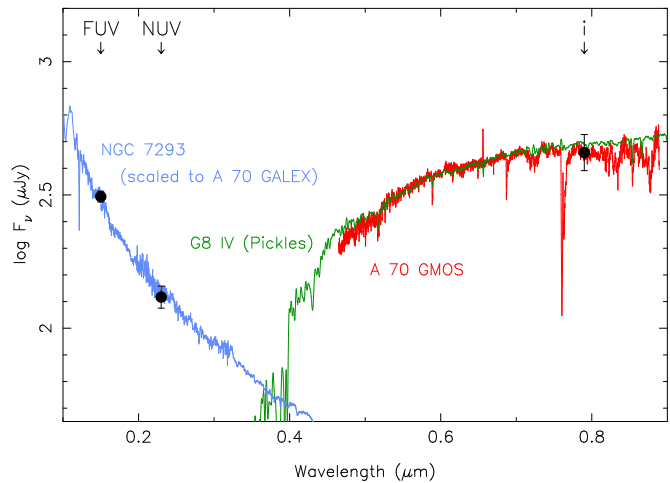}
   \end{center}
   \caption{Spectral energy distribution of the binary nucleus of A~70. The UV-excess of the hot component (represented by a scaled spectrum of NGC 7293) is clearly visible against the rapidly diminishing flux of the G8IV-V secondary towards UV wavelengths. Solid points mark fluxes obtained from photometric observations (see text).
   }  
   \label{fig:A70sed}
\end{figure}

To demonstrate that the \emph{GALEX} magnitudes belong to the WD we present the spectral energy distribution (SED) in Fig. \ref{fig:A70sed}. Fluxes were calculated using the relation $\mathrm{log}\ F_\nu[\mu\mathrm{Jy}] = -0.4m+9.56$.\footnote{http://galex.stsci.edu/GR4/?page=faq} for \emph{GALEX} magnitudes and standard flux at magnitude zero from Fouqu\'e et al. (2000) for the $i$ magnitude (Tab. \ref{tab:cool}). The cool component is shown by the dereddened GMOS R spectrum and is accompanied by a Pickles (1998) G8IV spectrum scaled to best fit the observed spectrum, while the hot component is the scaled NGC 7293 spectrum used to estimate the WD magnitudes. Note the clear UV-excess that confirms the presence of the hot WD primary to the s-process enhanced G8IV-V secondary that dominates at optical wavelengths. The intrinsic faintness of the WD ($V=20.4$ mag) and large magnitude difference ($\Delta V=2.6$ mag) explains why it was not seen in our spectroscopy.

\begin{table}
   \centering
   \caption{Observed ($m$) and dereddened ($m_0$) magnitudes of the WD component of the CSPN.}
   \begin{tabular}{crrrl}
      \hline\hline
      Waveband & $m$ & $m_0$ & Source \\
      \hline
      FUV & 18.0 & 17.6 & \emph{GALEX} \\
      NUV & 18.9 & 18.6 & \emph{GALEX} \\
      Johnson $U$  & 18.9 & 18.6 & scaled NGC 7293 \\
      Johnson $B$  & 20.2 & 19.9 & scaled NGC 7293 \\
      Johnson $V$  & 20.4 & 20.3 & scaled NGC 7293 \\
      Johnson $R$  & 20.5 & 20.4 & scaled NGC 7293 \\
      Johnson $I$  & 20.7 & 20.6 & scaled NGC 7293 \\
      \hline
   \end{tabular}
   \label{tab:hot}
\end{table}

\section{The Nebula}
\label{sec:nebula}
\subsection{Morphology}
\label{sec:imaging}
\begin{figure*}
   \begin{center}
      \includegraphics[scale=0.37]{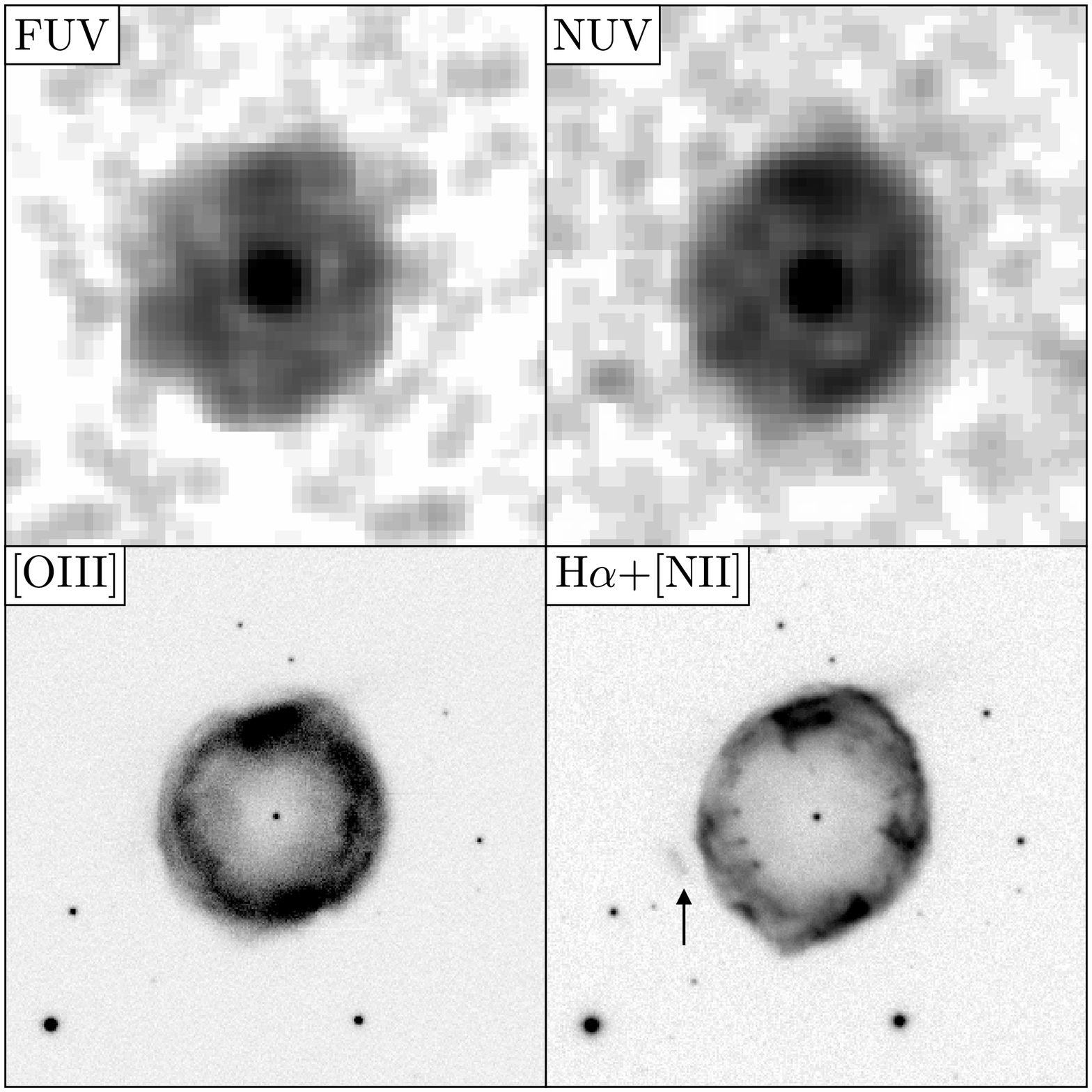}
      \includegraphics[scale=0.65]{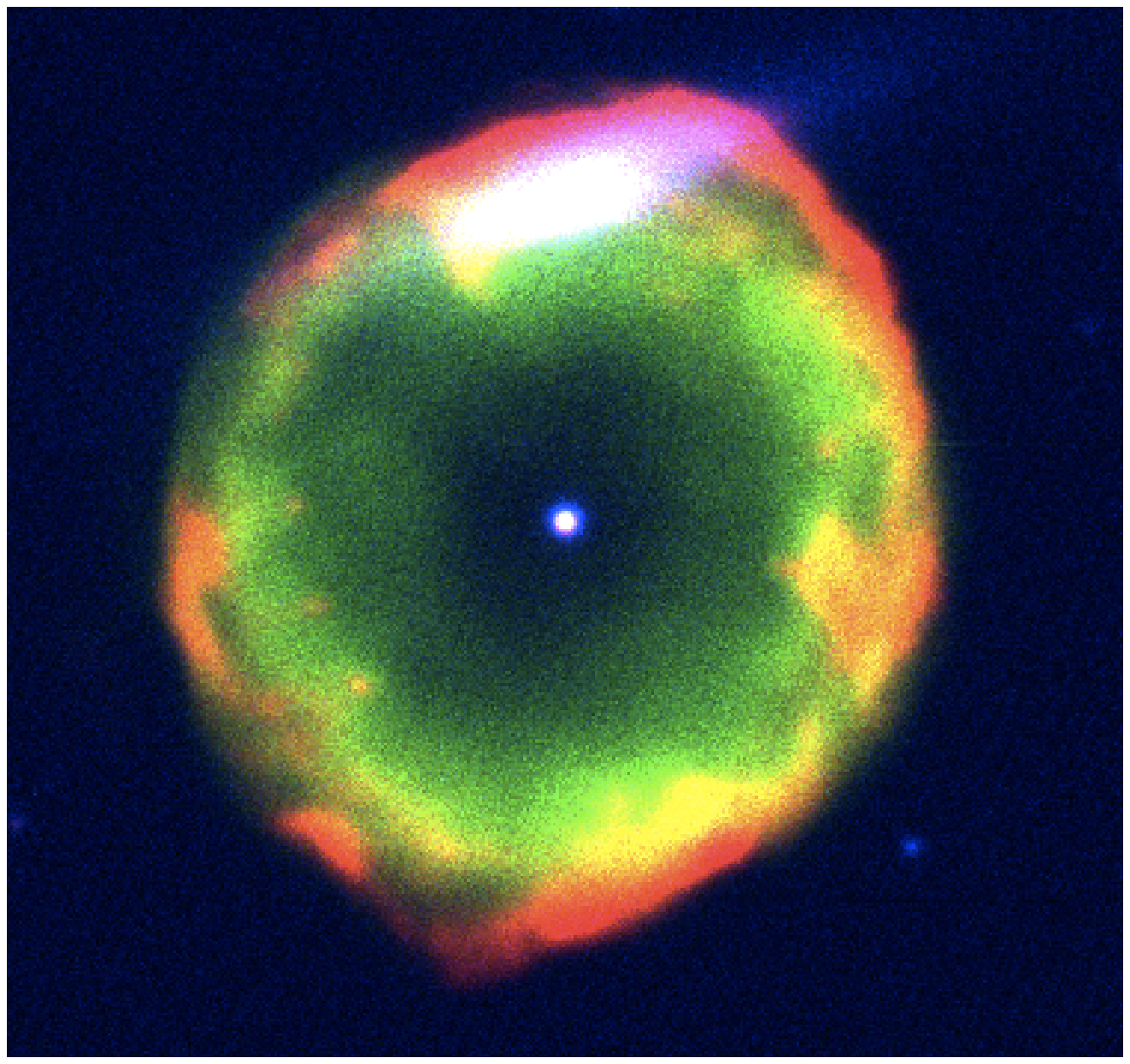}
   \end{center}
   \caption{A multi-wavelength view of A~70. \emph{(left)} \emph{GALEX} FUV and NUV images reveal the hot WD and nebula, while GMOS [O~III] and H$\alpha$+[N~II] images show the G8IV-V companion and the nebula in detail. Each image is 1.5$\times$1.5 arcmin$^2$ with North up and East to left and the external feature first noted by Hua et al. (1998) is arrowed. \emph{(right)} Colour-composite image made from GMOS HaC (red), OIII (green) and OIIIC (blue) exposures reveals the galaxy 6dFGS gJ203133.1-070502 superposed with the northern edge of the nebula creating a diamond ring effect.}
   \label{fig:a70}
\end{figure*}

GMOS acquisition images of 60 s each were taken in the OIII, OIIIC, Ha and HaC filters whose central wavelengths and FWHMs are 499.0/4.5 nm, 514.0/8.8 nm, 656.0/7.2 nm and 662.0/7.1 nm, respectively. An average stellar FWHM of $\sim$0.65\arcsec\, was measured from the images which are sampled at 0.145 \arcsec/pixel. No other stars besides the CSPN are detected within the nebula whose dimensions at 10\% of peak intensity are $45.2\times37.8$\arcsec (Tylenda et al. 2003). The Ha filter includes H$\alpha$ and both [N~II] lines, while the HaC filter is an [N~II] $\lambda$6584 filter. The [O~III] and H$\alpha$+[N~II] images are depicted in Fig \ref{fig:a70} together with the \emph{GALEX} FUV and NUV images whose 4--6\arcsec\ (FWHM) resolution was smoothed by a $2\times2$ pixel Gaussian.

The apparent morphology of A 70 is that of a ring nebula (e.g. NGC 6720). On closer inspection the [O~III] image shows a ridged appearance similar to Sp 1 which is a bipolar torus viewed close to pole-on (Bond \& Livio 1990; Jones et al. in prep.). Multiple knots of low-ionisation (Gon{\c c}alves et al. 2001) are also seen which are common in post-CE nebulae (Miszalski et al. 2009b). It is unlikely however that A~70 is the outcome of a CE interaction, although we cannot discard this possibility outright since the orbital period and eccentricity are not yet determined. While it is possible for the shortest period Ba stars to go through a CE phase, such occurrences are rare since the eccentricity distribution of Ba stars is explained by a combination of wind accretion and tidal evolution, rather than CE evolution or stable RLOF which tend to circularise orbits (Karakas, Tout \& Lattanzio 2000). 

The Barium star nature of A~70 tells us that wind interaction in a long orbital period binary has happened, and since bipolar nebulae can be produced under such circumstances (Mastrodemos \& Morris 1999; Gawryszczak, Miko{\l}ajewska \& R\'{o}\.{z}ycka 2002), we can conclude the nebula of A~70 may have been shaped in this fashion. The ring morphology of A~70 further strengthens the apparent trend seen already in similar PNe such as WeBo~1 (Bond et al. 2003), and the probably related Me~1-1 (Shen et al. 2004; Pereira et al. 2008), which both display bipolar morphologies. Pereira et al. (2008) suggest the barium abundance of Me~1-1 may have been diluted to explain the lack of barium enhancement, however its K(1-2) II nucleus does show a high rotation velocity of 90 km/s in common with other Barium stars.

Also seen in our images is the external feature outside the main nebula at position angle (PA) of 108$^\circ$, as first remarked by Hua et al. (1998), that is also visible in earlier images taken by Jewitt et al. (1986). The nature of this feature remains uncertain since its velocity (consistent with the main nebula) and [O~III]/H$\beta$ ratio (4.4) are both lower than would be expected for a shocked collimated outflow. It may be related to loop-like structures seen outside the main nebulae of bipolar PNe (e.g. K~3-17, Miranda, Ramos-Larios \& Guerrero 2010).

\subsection{Chemical abundances and plasma parameters}
\label{sec:chem}
\begin{figure*}
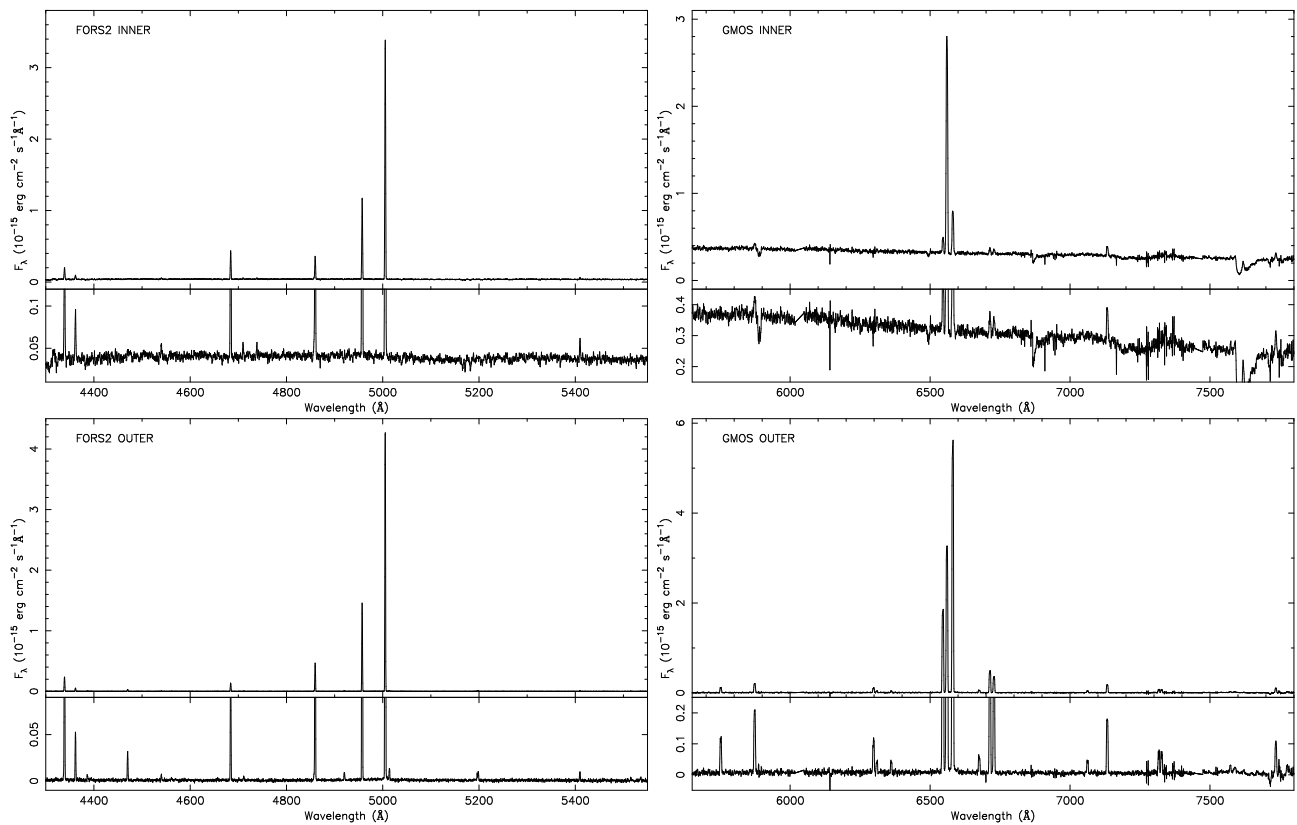

   \begin{center}
      \includegraphics[scale=0.375,angle=270]{blueinner.ps}
      \includegraphics[scale=0.375,angle=270]{redinner.ps}\\
     \includegraphics[scale=0.375,angle=270]{bluelow.ps}
     \includegraphics[scale=0.375,angle=270]{redlow.ps}\\
   \end{center}
   \caption{FORS2 (left column) and GMOS (right column) longslit spectroscopy of the inner zone and outer zone (see text).}
   \label{fig:nebula}
\end{figure*}

Earlier studies of the chemical properties of A~70 by Kaler et al. (1990) and Perinotto et al. (1994) showed an enhancement of helium. However, their spectra were taken only at bright parts of the nebula and were not deep enough to measure the weak diagnostic lines needed to derive accurate abundances. The greater depth of our spectra allows for a greatly improved abundance analysis which benefits from both [O~III] $\lambda$4363 and [N~II] $\lambda$5755 measured at S/N $\sim$20. We perform our analysis using the red GMOS spectrum and the FORS2 2010B spectrum which is slightly deeper than the blue GMOS spectrum and is taken at essentially the same PA (87$^\circ$ instead of 90$^\circ$). In order to measure the faintest lines in our spectra we separately extracted an \emph{inner zone}, from a 18.5" wide region centered on the central star that emphasises the highest ionisation species, and an \emph{outer zone}, being the average of two 9\arcsec\ zones either side of the inner zone extraction. Figure \ref{fig:nebula} shows the four extracted spectra where the bottom part of each panel highlights the faintest lines. Emission line intensities from these two extractions were combined to form an average spectrum for the whole nebula, such that the brighter emission line intensities matched those recovered from a separate extraction of the whole nebula. Table \ref{tab:lines} records our measurements for the inner zone and the combined average or total spectrum. The VLT and GMOS spectra were merged by rejecting lines bluer than $\lambda$5600 \AA\ in the GMOS spectrum and by scaling the spectra to obtain H$\alpha$/H$\beta$ = 3.0 (i.e. $c(H\beta)=0.07$, Acker et al. 1992). Line intensities were dereddened using the Howarth (1983) extinction law. An unidentified emission line at $\lambda$7738 \AA\ is associated with the lower ionisation O$^+$ region.

\begin{table}\centering
     \caption{The measured and dereddened emission line intensities.}
     \label{tab:lines}
     \begin{tabular}{lrrrr}
     \hline\hline 
                     & \multicolumn{2}{c}{Inner}& \multicolumn{2}{c}{Total} \\
      Identification & $F_\lambda$ & $I_\lambda$ & $F_\lambda$ & $I_\lambda$ \\
  \hline
  {}H I $4340$      & 51.6 & 52.7 & 50.4 & 51.4 \\
  {}[O III] $4363$  & 19.1 & 19.5 & 13.7 & 14.0 \\ 
  {}He I $4388$     &  -   &  -   &  1.0 &  1.0 \\
  {}He I $4472$     &  -   &  -   &  6.2 &  6.3 \\
  {}He II $4542$    &  5.1 &  5.2 &  3.2 &  3.2 \\
  {}He II $4686$    &  115.2 &  116.0 & 56.2 & 56.6 \\
  {}[Ar IV] $4711$  &  5.1 &  5.1 &  1.2 &  1.2 \\
  {}He I $4713$     &    - &  -   &  1.0 &  1.0 \\
  {}[Ne IV] $4720$  &  2.4 &  2.4 &  0.6 &  0.6 \\
  {}[Ar IV] $4740$  &  4.3 &  4.3 &  1.0 &  1.0 \\
  {}H I $4861$      &  100.0 &  100.0 &  100.0 &  100.0 \\
  {}He I $4922$     &    - &    - &  1.8 &  1.8 \\
  {}[O III] $4959$  &  349.2 &  347.9 &  336.2 &  334.9 \\
  {}[O III] $5007$  & 1036.9 & 1031.0 &  992.8 &  987.1 \\
  {}He I $5015$     &    - &    - &  2.7 &  2.6 \\
  {}[N I] $5200$    &    - &    - &  4.8 &  4.8 \\
  {}He II $5412$    &  8.1 &  7.9 &  3.9 &  3.8 \\
  {}[Cl III] $5518$ &    - &    - &  0.4 &  0.4 \\
  {}[Cl III] $5538$ &    - &    - &  0.4 &  0.4 \\
  {}[N II] $5754$   &    - &    - & 14.3 & 13.9 \\
  {}He I $5876$     &  7.9 &  7.6 & 19.4 & 18.7 \\
  {}[O I] $6300$    &    - &    - & 13.7 & 13.1 \\
  {}[S III] $6312$  &    - &    - &  3.4 &  3.2 \\
  {}[O I] $6364$    &    - &    - &  5.4 &  5.1 \\
  {}[N II] $6548$   & 19.0 & 18.0 &  221.8 &  210.7 \\
  {}H I $6563$      &  300.0 &  284.9 &  300.0 &  284.9 \\
  {}[N II] $6583$   & 59.8 & 56.8 &  673.5 &  639.4 \\
  {}He I $6678$     &    - &    - &  5.7 &  5.4 \\
  {}[S II] $6716$   &  7.4 &  7.0 & 60.6 & 57.3 \\
  {}[S II] $6731$   &  6.7 &  6.3 & 42.7 & 40.4 \\
  {}He I $7065$     &    - &    - & 4.3  &  4.0 \\
  {}[Ar III] $7136$ & 13.3 & 12.5 & 17.3 & 16.2 \\
  {}[O II] $7320$   &    - &    - &  3.8 &  3.6 \\
  {}[O II] $7330$   &    - &    - &  3.8 &  3.6 \\
  {}?? $7738$       &    - &    - &  5.5 &  5.1 \\
  \hline
     \end{tabular}
  \end{table}

The measured line intensities were analysed with the plasma diagnostics program HOPPLA (Acker at al. 1991, K\"oppen et al. 1991; see also Girard et al. 2007). 
Electron temperatures in the O$^{+}$ and O$^{++}$ zones
were derived from the [N~II] and [O~III] line ratios 
in a consistent way with the electron density (see Table \ref{tab:plasma}). 
The [S~II] line ratio is in its low-density limit, while both [Cl~III] 
and [Ar~IV] yielded substantially higher densities. 
We have therefore determined the chemical composition for two density values in Tab. \ref{tab:abundances} where the usual ionisation correction factors yield the elemental abundances expressed in the usual 12+log($n($X$)/n($H$)$) format. Table \ref{tab:abundances} also includes our HOPPLA reanalysis of Perinotto et al. (1994) and Kaler et al (1990),\footnote{Note the previously published line intensities for A~70 do not include the weak diagnostic and auroral lines required to derive meaningful abundances.} the average values for Type-I and non Type-I PNe from Kingsburgh \& Barlow (1994), and the solar abundances from Asplund et al. (2005). Values in parentheses are results without proper correction for unseen ionic stages, and should be considered strictly as lower limits, while colons in Tab. \ref{tab:abundances} indicate quantities with greater uncertainty.

  \begin{table}\centering
     \caption{The derived plasma parameters.}
     \label{tab:plasma}
     \begin{tabular}{lrr}
  \hline \hline
       Parameter & Inner & Total \\
    \hline
      $c(H\beta)$ (Acker et al. 1992)   & 0.07  &  0.07  \\
      $T_e$([O III]) [K]          & 14900 &  13200 \\
      $T_e$([N II]) [K]           &   - &  12400 \\
      $n_e$([S II]) [cm$^{-3}$]   &   300 & $\le$ 100 \\
      $n_e$([Cl III]) [cm$^{-3}$] &   - &  2700  \\
      $n_e$([Ar IV]) [cm$^{-3}$]  &  1700 &  1600  \\
     \hline
     \end{tabular}
  \end{table}

  \begin{table*}
     \centering
     \caption{Chemical abundances and plasma parameters of A~70.}
     \label{tab:abundances}
     \begin{tabular}{lrrrrrrrr}
      \hline \hline 
         & Inner & \multicolumn{2}{c}{Total} & KSK1990 & P1994 &Type I & non-Type I & Sun\\
     \hline
    $n_e$ [cm$^{-3}$] & 301   & 100   & 3000  &    40   &  2000 &-&-&-\\
    $T_e$(O$^+$) [K]     & 8900: & 12400 & 11800 & 10200   & 15000 &-&-&-\\
    $T_e$(O$^{++}$) [K] & 14900 & 13200 & 13100 &  9700   & 15000 &-&-&-\\
    $T_e$(He$^{++}$) [K] & 14900 & 12800 & 12400 &  9700   & 15000 &-&-&-\\
      \hline
      He               & 11.19 & 11.28 & 11.23 & -    & 11.26  & 11.11 & 11.05 & 10.93 \\
      N                & (6.97)&  8.68 &  8.96 &$>8.41$ &$>7.52$ &  8.72 &  8.14 &  7.78 \\
      O                &  8.51 &  8.43 &  8.42 & 8.68   &  7.98  &  8.65 &  8.69 &  8.66 \\
      Ne               & 7.84::  & (7.68)& (7.71)& -    & -    &  8.09 &  8.10 &  7.84 \\
      S                & (5.44)&  6.82 &  7.04 & -    & -    &  6.91 &  6.91 &  7.14 \\
      Cl               &  -  &  4.61 &  4.69 & -    & -    &      - &-       &   5.5 \\
      Ar               & (5.87)&  6.11 &  6.25 & -    & -    &  6.42 &  6.38 &  6.18 \\
     \hline
      log(N/O)         &  -  & +0.25 & +0.54 &$>-0.27$&$>-0.34$&$+0.07$&$-0.55$&$-0.88$\\
      log(Ne/O)        &$-0.58$&  -  &  -  & -    & -    &$-0.56$&$-0.59$&$-0.82$\\
      log(S/O)         &  -  &$-1.61$&$-1.38$& -    & -    &$-1.74$&$-1.78$&$-1.52$\\
      log(Ar/O)        &  -  &$-2.32$&$-2.17$& -    & -    &$-2.23$&$-2.31$&$-2.28$\\
     \hline
     \end{tabular}
  \end{table*}

Due to the lower critical density collisional de-excitation for the [N~II] lines, the two assumptions for the electron density affect only the nitrogen abundance. In any case, the He/H and N/O ratios both exhibit a strong enhancement which, together with the overall abundance pattern, makes A~70 a genuine Type-I object (Peimbert \& Torres-Peimbert 1983; Kingsburgh \& Barlow 1994). As highly evolved nebulae are expected to be low density objects, we attach
a greater importance to the analysis made with $n_e = 100$ cm$^{-3}$. It may well
be that [Ar~IV] and [Cl~III] emission originates in denser clumps, but in either
analysis the contribution of Ar$^{3+}$ to the elemental abundance is only 
10 percent. Hence, the argon abundance is essentially determined by [Ar~III].
Similarly, sulphur is derived in almost equal terms from [S~II] and [S~III],
for which the low density deduced from the [S~II] line ratio is more representative.  

To estimate the errors in our abundance analysis we performed Monte Carlo simulations. This involved 100 iterations of adding wavelength-independent noise to our spectra and reperforming our analysis each time. The noise level selected was 0.5\% of H$\beta$ as judged from measurement of our spectra. From the standard deviation of values derived in our simulations we find an error of 200 K for $T_e$([O~III]) and $T_e$([N~II]), 0.009 dex for the He abundance, 0.018 dex for O, 0.034 dex for N, 0.028 dex for Ar and 0.049 dex for S. These are however formal errors since uncertainties concerning the ionisation correction factor and `averaging' $T_e$([O~III]) give a lower limit to the actual error of at least 0.1 dex for the oxygen abundance. Since weak emission lines were used to determine the Ne and Cl abundances, and the [Ar~IV] and [Cl~III] densities, these values may only be seen as indicative. Since the spectra did not cover the [Ne~III] lines, we could only estimate the neon abundance from the [Ne~IV]/H$\beta$ ratio obtained from our extraction of the inner nebula. Thus, the Ne/O abundance ratio would be at least $-0.58$, which is larger than in the Sun, but close to the average found in PNe. 

A number of consistency checks were also applied to the total spectrum.
The intensities of the three Balmer lines match within 8 percent and we checked whether all lines of the same ion should give the same ionic abundance. The He~I 
lines give the same ionic abundance within about 10 percent, if one 
excludes $\lambda$5015 \AA\ which is underestimated by a factor of 1.4, 
and the very weak lines $\lambda$4713 and $\lambda$4388 \AA. Among the 
three He~II lines $\lambda$4541 \AA\ is overestimated by a factor of 1.7. 
The three lines of [N~II] and [O~III] have the same ionic abundances within 3 percent.

\subsection{Distance}
\label{sec:distance}
Table \ref{tab:dist} lists two distance estimates of 2.4 kpc (Stanghellini, Shaw \& Villaver  2008, hereafter SSV08) and 5.0 kpc from the mean-trend of the H$\alpha$ surface brightness-radius relation (SBR) of Frew \& Parker (2006). The SBR calculation adopted an integrated H$\alpha$ flux of log $F(H\alpha)=-11.85$, the mean of fluxes measured by Kaler (1983) and Hua et al. (1998), angular dimensions from Tylenda et al. (2003), and the reddening from Acker et al. (1992). Also given in Tab. \ref{tab:dist} are distance-dependent luminosities of the central star components, the nebula ionised mass $M_\mathrm{ion}$ (assuming a filling factor of 0.3), the height below the Galactic Plane $z$ and the expansion age $t_\mathrm{exp}$ of the nebula (using 40 km/s measured by Meatheringham et al. 2008).

\begin{table}
   \centering
   \caption{Distance estimates for A~70 and distance-dependent quantities.}
   \label{tab:dist}
   \begin{tabular}{cccccrc}
      \hline
      \hline
      $d$ & $M_V$ & $M_V$ & $M_\mathrm{ion}$ & $z$ &$t_\mathrm{exp}$ &  Method\\
      (kpc) & cool & hot & ($M_\odot$) & (kpc) & (yrs) & \\
      \hline
      2.4 &   5.8       &  8.4          &   0.05     &  -1.0               &  5700     & SSV08 \\
      5.0 &   4.2       &  6.8          &   0.29     &  -2.1               &  12000    & SBR \\
      \hline
   \end{tabular}
\end{table}

Further distances imply a higher $z$, higher $M_\mathrm{ion}$, higher stellar luminosities and older $t_\mathrm{exp}$, while shorter distances imply a lower $z$, lower $M_\mathrm{ion}$, lower stellar luminosities and a younger $t_\mathrm{exp}$. If the SSV08 distance were adopted the stellar luminosities would imply an implausibly massive ionising star for the small associated $t_\mathrm{exp}$ and $M_\mathrm{ion}$ would be unusually low for a Type I PN ($M_\mathrm{ion}$ can reach 0.5--1 $M_\odot$, see Frew, Parker \& Russeil 2006). For these reasons we favour the 5.0 kpc SBR distance, although the accompanying $z$ is difficult to reconcile with the only moderately sub-solar oxygen abundance of 8.4 dex ($-0.3$ dex c.f. solar). A shorter distance would alleviate but not solve this problem, although a Thick Disk progenitor may be one solution to the problem. 

The Type-I abundance of A~70 also places it against the general trend of decreasing N/O with increasing $z$ (K\"oppen \& Cuisinier 1997) and the empirical finding that most bipolar Type-I PNe are found at very low $z$ (Corradi \& Schwarz 1995). Whether the binary nature of A~70 could explain this incongruity remains to be proven. Irrespective of the adopted distance, we can say that the white dwarf has turned onto the WD cooling track (e.g. Frew et al. 2006) and that the luminosity of the companion is less certain. It could be either G8V (for shorter distances) or G8IV (for larger distances, see Sandage et al. 2003). Improved central star magnitudes and spectroscopy are required to decide between each luminosity class, so for now we adopt G8IV-V.

\section{Discussion}
\subsection{Evolutionary status} 
The G8IV-V star with Ba~II and Sr~II enhancement, chromospheric H$\alpha$ emission and negligible to low radial velocity amplitude, coupled with the \emph{GALEX} UV detection of the WD, all strongly support the presence of a Barium star nucleus in A~70. Having detected a PN around a Barium star is quite rare, however A~70 may represent an even more transient stage of stellar evolution if the secondary can be proven to be a sub-giant (Sandage et al. 2003). The evolutionary history of A~70 starts with the slightly more massive component of a binary system with a mass ratio near unity evolving through the AGB, and after experiencing thermal pulses, becomes s-process enhanced. Through heavy mass loss its s-process rich wind contaminated its less massive companion and reversed the mass ratio of the system to produce a Barium star. 
The AGB star then reached the PN phase, while its companion left the main sequence to become a subgiant, at which point we observe A~70 -- the penultimate phase in the making of a Barium star! A~70 thus represents an evolutionary stage just before WeBo~1 (Bond et al. 2003), which is a carbon and s-process enhanced late-type giant star surrounded by a ring-like planetary nebula. Similar to A~70, WeBo~1 is chromospherically active and it undergoes rapid rotation every 4.7 days. Such a short period, which is not expected as tidal forces should have synchronised the stars, may be due to spin-up of material accreting from the companion (Theuns et al. 1996; Jeffries \& Stevens 1996). 

Given the present state of the system, we can also consider how A~70 will evolve, although the final state will depend on many parameters. In a few thousand years the nebula will disappear and the hot star will gradually cool off. Soon after this the cool star will ascend the red giant branch (RGB) and the system will appear as a common Barium star with a peculiar red giant and no visible companion. 
At the tip of the RGB the star will experience heavy mass loss by transferring matter to the white dwarf. During this time it may be observed as a symbiotic system where accreting matter will heat the white dwarf and ionise the wind of the cool giant. 

The Barium stars have close ties with symbiotic stars as well as with planetary nebulae. Several extrinsic S stars have been found to exhibit symbiotic-like features (e.g. Ake et al. 1991), while several symbiotic stars are also known to present overabundances of s-process elements (Smith et al. 1996, 1997; Pereira et al. 2005). The D'-type symbiotic stars are the closest relatives to Barium stars (Schmid \& Nussbaumer 1993; Pereira et al. 2005; Jorissen et al. 2005), especially considering some of them are surrounded by apparently \emph{planetary} nebulae. Schwarz (1991) discovered an inner and outer nebula around one of these, AS~201, of which the outer nebula is likely to be a PN (ejected by the WD dwarf). Miszalski et al. (2011c) found symbiotic characteristics in the Galactic Bulge PN M~2-29, which also exhibits inner and outer nebulae, however the secondary has yet to observed against the glare of the primary. Appendix \ref{sec:app} describes the discovery of a bipolar nebula around HD~330036 (Cn~1-1), which if considered to be a \emph{planetary} nebula, would add further evidence to the link between Barium stars with PNe and barium enhanced D'-type symbiotic stars. 

Observing a system such as A~70 implies that the initial mass ratio must have been close to unity. The rarity of such a configuration may be used as an argument against the present formation scenario, and indeed an identical approach was taken by Corradi (2003) concerning the potential presence of Mira secondaries in PNe. Such an A~70-like configuration can in principle occur since Lucy (2006) found an excess of so-called twins, i.e. systems with mass ratios between 0.98 and 1. As further binaries similar to A~70 and WeBo~1 are found, then these probabilities may have to be revised in favour of a greater frequency of twins. This would foster a greater overlap between symbiotic stars (at least those of yellow or D'-type) and PNe (Jorissen et al. 2005), with the formal difference becoming notional. Wide binaries will interact and produce genuine PNe in a wide variety of cases, however only a relatively narrow range is currently observed (e.g. De Marco 2009).

\subsection{A~70 as a probe of AGB nucleosynthesis}
The measurement of nebular s-process abundances in PNe has considerable potential to improve AGB nucleosynthesis models (e.g. Sterling \& Dinerstein 2008, hereafter SD08; Karakas et al. 2009; Karakas \& Lugaro 2010). These abundances are a valuable constraint upon the number of third dredge-up episodes experienced during the thermally-pulsing AGB phase. Of particular interest are Type-I PNe whose He- and N-rich abundances are well-reproduced in models that require a progenitor mass $>4$ M$_\odot$ to achieve hot bottom burning. Quantifying the stellar s-process abundances of A~70 via high-resolution spectroscopy will therefore be of great interest to compare against the Type-I nebula abundance. The high surface brightness of A~70 compared to e.g. WeBo~1 may also allow nebula s-process abundances to be measured via NIR spectroscopy. 

In principle the s-process abundances of Type-I PNe should be relatively straight-forward to understand, however no firm patterns have been found so far (Karakas et al. 2009). SD08 noted that binary interactions may be responsible for reducing their measured s-process abundances, however the paucity of known binaries in their sample meant this remained untested. New discoveries of binary central stars in the SD08 sample should therefore help resolve the issue. Miszalski et al. (2011b) recently found a close binary in NGC~6778 which is a bipolar Type-I PN in the SD08 sample with strong He and N enhancement (He/H=0.155 and N/O=$+$0.78, Perinotto, Morbidelli \& Scatarzi 2004). It is possible that post-CE binaries may reduce s-process abundances to a greater extent than wider binary systems, making wider binaries a potentially more reliable probe of s-process abundances in PNe. The growing sample of A~70 and WeBo~1 as Barium stars, and perhaps also Me~1-1, therefore provides a powerful alternative to analysis of close binaries in the SD08 sample. 

\subsection{Proving a PN has a cool companion}
The discovery of the cool CSPN of A~70 has important implications for further survey work to find similar binaries. To establish a physical connection between a cool CSPN candidate and a PN a number of factors must be considered. The two most important being: (i) a UV excess or spectroscopic features of the WD must be detected to rule out the born-again scenario, and (ii) agreement between the radial velocities of the nebula and the cool star (modulo the expected radial velocity amplitude of a binary). It is also preferable to have a small nebular diameter or equivalently an uncrowded field to help rule out a superposition (e.g. Ciardullo et al. 1999). A favourable spectroscopic distance should also preferably agree well with the approximate statistical nebula distance. 

A general lack of deep enough UV observations has made the task of confirming hot components in PNe difficult. Sahai et al. (2008) found 9/21 AGB stars to have FUV excesses which they attribute to a hot companion, but there has not been a systematic UV-excess survey of PNe. The only other object studied so far is K~1-6 (Frew et al. 2011), however no spectroscopic data were presented in this study. We also advise against searching too hard for a cool companion, especially in large PNe where many candidates may be present. It is certainly possible that an intrinsically very faint $M_V\ga+7$ mag single WD nucleus may be beyond the detection limit of typical optical surveys, and this is especially true for the most distant PNe. This emphasises the crucial role of UV photometry to confirm such A~70-like binaries if they were located in more crowded stellar fields. On the other hand, the unique spectroscopic signature of a Barium star is also a sound means to secure a binary confirmation and multi-object spectroscopy may be an efficient tool in this respect.

\section{Conclusions}
\label{sec:conclusion}
We have presented 8-m optical spectroscopic and imaging observations of the unique diamond-ring PN A~70. Combined with \emph{GALEX} UV photometry, the data prove the binary nature of the central star which is expected to have an expected orbital period of a few hundred days. Our main conclusions are as follows:

\begin{itemize}
   \item Optical spectroscopy of the central star revealed a G8IV-V star, slightly enhanced in s-process elements, with chromospheric H$\alpha$ emission. \emph{GALEX} UV photometry detected the signature of a WD coincident with the central star, therefore providing the first clear proof for a Barium dwarf binary inside a PN. This is firm evidence for the standard formation scenario of Barium stars since Barium dwarfs are generally difficult to observe (e.g. Gray et al. 2011).
   \item Radial velocities of the G8IV-V secondary were analysed in comparison to the $-72\pm3$ km/s heliocentric nebula velocity of A~70. FORS2 observations are consistent with the nebula velocity proving they are physically connected, while one GMOS observation presented a difference of 24 km/s. Such a difference is of the same magnitude expected for orbital motion, if indeed this one measurement can be trusted. Further observations are required at higher resolution to measure the orbital period and RV amplitude.
   \item Chemical abundances of the nebula were measured and found to have a Type-I composition with strong He and N enrichment. As the G8IV-V companion is also s-process enhanced, this makes A~70 an exceptional laboratory for further improving our understanding of AGB nucleosynthesis and the origin of Type-I PNe (e.g. Karakas et al. 2009). 
   \item The distance remains uncertain, however it is clear that the WD has evolved onto the cooling track. A plausible 5 kpc distance would imply a height below the Galactic Plane of $z\sim-2.0$ kpc, in contradiction with the Type-I composition and the only slightly sub-solar oxygen abundance ($-0.3$ dex c.f. solar). It may be possible that a Thick Disk origin or binary stellar evolution could explain the Type-I composition, but this remains to be proven with improved observations. 
   \item A~70 is one of very few PNe with apparently cool central stars to have the WD detected in the UV. The observations presented here strongly suggest that insensitive UV observations were responsible for the non-detections of Bond \& Pollacco (2002) and the conclusion that the companions are single nuclei caught during the `born-again' phase. A~70 also serves as a template to guide future discoveries of similar binary central stars. 
\end{itemize}

\section*{Acknowledgments}
  We acknowledge the conscientious and helpful support of Simon O'Toole and Steve Margheim during the course of our Gemini program.
   BM thanks Observatoire astronomique de Strasbourg for travel support towards the later stages of this work and R. Napiwotzki for assistance in building the SED of A~70.
    HB thanks Leo Rivas and Yuri Beletsky for their dedication and excellence at the VLT. Their persistence in staying until the last moments of the night in the control room is especially appreciated. 
   AFJM is grateful to NSERC (Canada) and FQRNT (Quebec) for financial assistance.
    SYNPHOT is a product of the Space Telescope Science Institute, which is operated by AURA for NASA.
    Based on observations obtained at the Gemini Observatory, which is operated by the
    Association of Universities for Research in Astronomy, Inc., under a cooperative agreement
    with the NSF on behalf of the Gemini partnership: the National Science Foundation (United
    States), the Science and Technology Facilities Council (United Kingdom), the
    National Research Council (Canada), CONICYT (Chile), the Australian Research Council
    (Australia), Minist\'erio da Ci\^encia e Tecnologia (Brazil) and Ministerio de Ciencia, Tecnolog\'ia e Innovaci\'on Productiva (Argentina).

\appendix

\section[]{The bipolar nebula of HD~330036}

\label{sec:app}
\begin{figure}
   \begin{center}
      \includegraphics[scale=0.5]{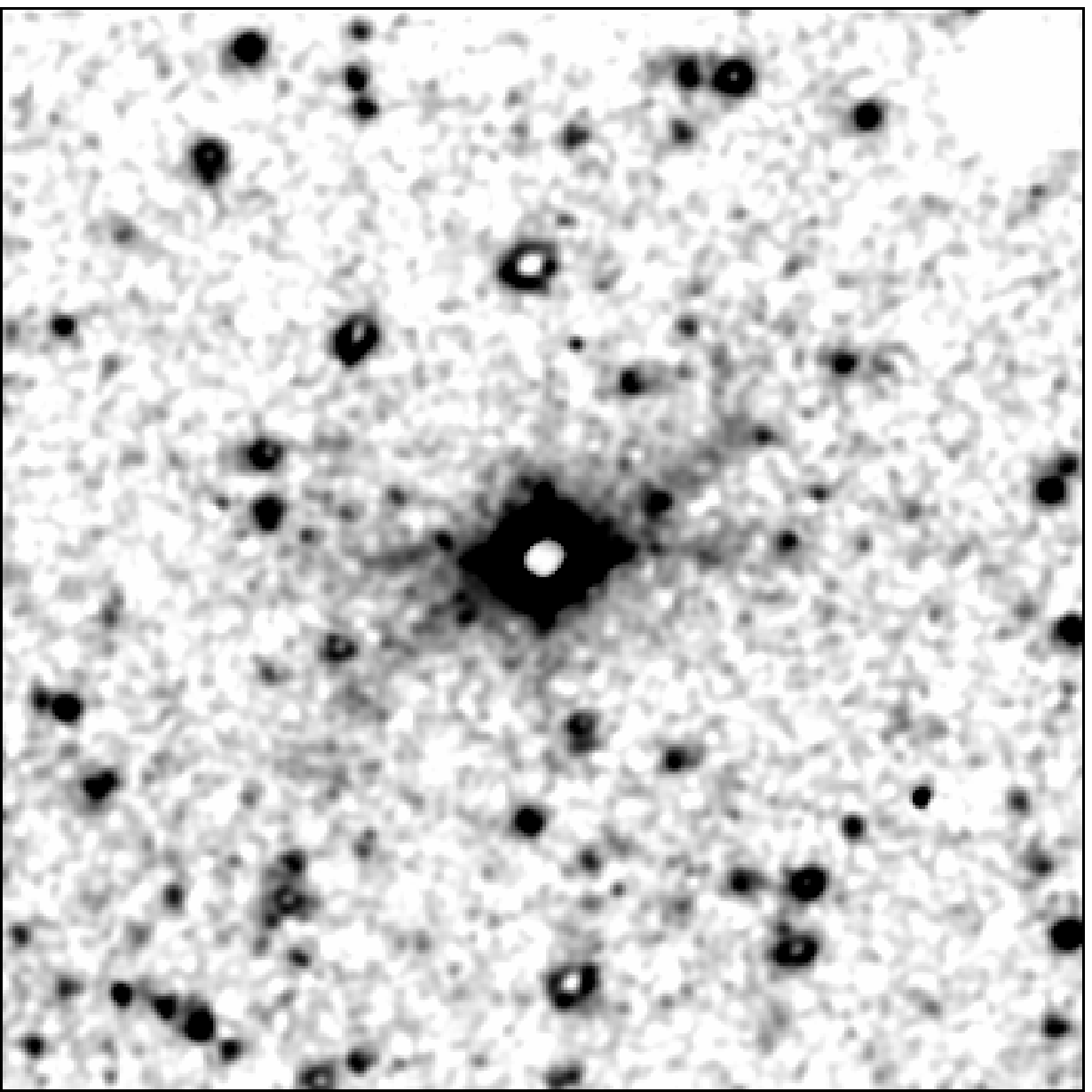}
   \end{center}
   \caption{A pair of faint bipolar or $X$-shaped lobes are seen around HD~330036 (Cn~1-1) in this $4\times4$ arcmin$^2$ continuum-divided H$\alpha$ image (North is up and East to left). The opposing parts of the lobes align along PA$\sim$$90^\circ$ and PA$\sim$$125^\circ$ and measure 105--110\arcsec\ tip-to-tip.} 
   \label{fig:bipolar}
\end{figure}

The AAO/UKST SuperCOSMOS H$\alpha$ Survey (SHS, Parker et al. 2005) is a deep 4000 deg$^2$ H$\alpha$ and broadband red (Short-Red) photographic survey of the Southern Galactic Plane. Figure \ref{fig:bipolar} depicts a faint bipolar nebula in the SHS data surrounding HD~330036 (PN G330.7$+$04.1, also known as Cn~1-1) found during the course of this work. This is an entirely new discovery since previous studies were concerned with the bright emission-line core only and were not sensitive enough to reveal the very low surface brightness lobes (e.g. Kohoutek 1997). The flattened $X$-shape is uncannily similar to the `outer lobes' of Hen~2-104 (Corradi 2003), suggesting a similar process was responsible for their formation. At the 2.3 kpc distance estimated by Pereira et al. (2005) the lobes would measure 1.2 pc tip-to-tip.

\label{lastpage}

\end{document}